\documentclass[twocolumn,preprintnumbers,amsmath,amssymb]{revtex4}
\usepackage{graphicx}
\usepackage{dcolumn}
\usepackage{bm}

\begin{document}


\title{Asymmetric diffusion at the interfaces in multilayers }

\author{Ajay Gupta }
\email {agupta@csr.ernet.in}
\author {Dileep Kumar, Vaishali Phatak }
\affiliation {UGC-DAE Consortium for Scientific Research,
University Campus, Khandwa Road, Indore-452017, India\\}
\date{26 September 2009}

\begin{abstract}

Nanoscale diffusion at the interfaces in multilayers plays a vital
role in controlling their physical properties for a variety of
applications.  In the present work depth-dependent interdiffusion
in a Si/Fe/Si trilayer has been studied with sub-nanometer depth
resolution, using x ray standing waves.   High depth-selectivity
of the present technique allows one to measure diffusion at the
two interfaces of Fe namely, Fe-on-Si and Si-on-Fe, independently,
yielding an intriguing result that Fe diffusivity at the two
interfaces is not symmetric.  It is faster at the Fe-on-Si
interface.  While the values of activation energy at the two
interfaces are comparable, the main difference is found in the
pre-exponent factor suggesting different mechanisms of diffusion
at the two interfaces.  This apparently counter-intuitive result
has been understood in terms of an asymmetric structure of the
interfaces as revealed by depth selective conversion electron
M\"{o}ssbauer spectroscopy.  A difference in the surface free
energies of Fe and Si can lead to such differences in the
structure of the two interfaces.

\end{abstract}

\maketitle


Atomic diffusion is fundamental to many processes in material
science such as microstructure development, non-martensitic phase
transformation, stress relaxation etc.  In multilayers, atomic
diffusion at the interfaces plays a vital role in controlling
their physical properties for a wide variety of applications.  In
x ray and neutron mirrors intentionally diffused interfaces have
been used to reduce the higher order
contamination~\cite{Padiyath:APL:2006}.  In tunnel
magnetoresistance multilayers, thermal annealing can increase
magnetoresistance by orders of magnitude~\cite{Scola:APL:2006}. In
giant magnetoresistance (GMR) multilayers, interdiffusion can
significantly affect the GMR~\cite{Iusan:PRB:2007}.  In spin
valves with Mn based antiferromagnetic layer, Mn diffusion can
seriously degrade the performance~\cite{Jang:APL:2002}.  Co
diffusion in Sm-Co/Fe exchange-spring magnet films is known to
improve the exchange coupling~\cite{Choi:PRB:2007}.  While atomic
diffusion in bulk solids is a widely studied and fairly well
understood phenomenon, a reasonable understanding of the
interfacial diffusion in multilayers has yet to emerge.  Several
factors like a steep concentration gradient at the interfaces,
interfacial stresses and disorder may significantly modify the
diffusion in multilayers. This has resulted in unexpected
interfacial phenomenon like a non-parabolic shift of phase
boundaries in the presence of strong composition dependence of
diffusivity~\cite{Z.Erdelyi:Science:2004}.

X ray standing waves generated by total external reflection of x
rays from buffer layer~\cite{A.Gupta:PRB:2005}, or in a
multilayer~\cite{Bedzyk:Science:1990,Gupta:PRB:2007} have been
used for concentration profiling of various elements.  In the
present work, we exploit the depth selectivity of x ray standing
waves for studying the depth dependent interdiffusion in a
Si/Fe/Si trilayer with sub-nanometer depth resolution.  High depth
sensitivity of the technique allows one to measure diffusion at
the two interfaces of Fe namely Fe-on-Si and Si-on-Fe
independently.  An intriguing finding of the present work is that
the Fe diffusivity at the two interfaces is not symmetric;
diffusion is faster at the Fe-on-Si interface.  In order to
understand this apparently counter-intuitive result, conversion
electron M\"{o}ssbauer spectroscopy (CEMS) has been used to study
the interfacial structure. It is found that there is a significant
difference in the structure of the two interfaces, resulting in
different diffusivities at the two interfaces.  A difference in
the surface free energies of Fe and Si can lead to such difference
in the structure of the two interfaces.

\begin{figure} \center
\vspace {-0.5cm}
\includegraphics [width=50.73mm,height=56mm]{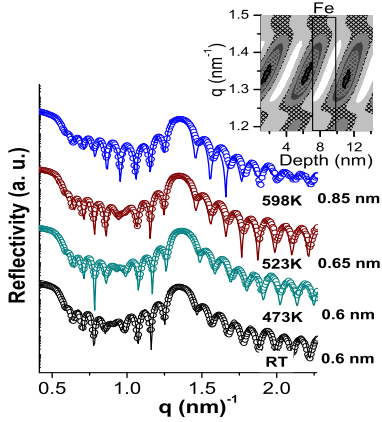}
\vspace {-0.3cm} \caption{\label{fig:fig1} Representative XRR of
SW$_{-}$ML after various stages of annealing. The continuous
curves represent the best fit to the experimental data. The
numbers shown against each curve represent annealing temperature
and the average interface roughness of W/Si multilayer
respectively. Inset shows the contour plot of x ray intensity as a
function of q and depth from the surface of the multilayer. The
rectangle represents the position of Fe layer.}
\end{figure}

The structure of the multilayer used for diffusion measurements is
[W (2.0\,nm)/Si (3.1\,nm)]$_{10}$ / Si (3.8\,nm)/ Fe (2.7\,nm)/Si
(7.0\,nm) (referred as SW$_{-}$ML).  The deposition was done using
ion beam sputtering in a vacuum chamber with a base pressure of
1$\times$10$^{-7}$\,mbar~\cite{Gupta:ASS:2003}.  A broad beam
Kaufman type ion source was used with Ar ions of energy 1\,keV and
a beam current of 20\,mA. The bottom [W (2.0\,nm)/Si
(3.1\,nm)]$_{10}$ multilayer is used to generate x ray standing
waves~\cite{Gupta:PRB:2007}.  On the top of this multilayer Si
(3.8\,nm)/Fe (2.7\,nm)/Si (7.0\,nm) structure was deposited
without breaking vacuum.  A thickness of 3.8\,nm for the first Si
layer is chosen in such a way that the Fe layer lies roughly
midway between two antinodes of the x ray standing waves generated
in W/Si multilayer at the Bragg peak. This point is clear from the
inset of Fig.~\ref{fig:fig1} which shows the contour plot of x ray
intensity as a function of depth and the scattering vector $q$. At
$q = 1.35\,nm^{-1}$ which corresponds to the center of the Bragg
peak of W/Si multilayer, Fe layer is midway between the two
antinodes.

Simultaneous x ray reflectivity (XRR) and x ray fluorescence (XRF)
measurements were done using Bruker D8 diffractometer fitted with
a G\"{o}bble mirror on the incident beam side in order to obtain a
parallel monochromatic beam of Cu K$\alpha$ radiation.
Fluorescence spectrum was measured using a Ketek detector with an
energy resolution of 200\,eV.  Thermal annealing of multilayer
structure was done in a vacuum of 10$^{-6}$\,mbar in order to
induce interdiffusion at the interfaces of Fe and Si layers.  A
separate study on the annealing behavior of W/Si multilayer showed
that at least up to 623\,K, the W/Si multilayer structure is
stable with almost no change in its x ray reflectivity.  Thus the
thermal annealing in the present multilayer essentially induces
interdiffusion at the interfaces of Fe and Si layers only.

In order to study the Fe-on-Si and Si-on-Fe interfaces using CEMS,
the following two multilayer structures were also prepared using
ion beam sputtering: (i) substrate/
[Si(6\,nm)/$^{57}$Fe(2\,nm)/Fe(2\,nm)]$_{10}$, and (ii) substrate/
[Si(6\,nm)/Fe(2\,nm)/ $^{57}$Fe(2\,nm)]$_{10}$, designed as ML1
and ML2 respectively.  In the first multilayer the 2\,nm thick Fe
layer lies on Fe-on-Si interface while in the second multilayer it
lies on Si-on-Fe interface.  Since M\"{o}ssbauer measurements are
sensitive only to $^{57}$Fe, the first multilayer gives
information preferentially about the Fe-on-Si interface, while the
second one gives information about Si-on-Fe interface.  The CEMS
measurements were done using a Wissel M\"{o}ssbauer spectrometer
and a gas flow proportional counter with 95\% He + 5\% CH$_{4}$.


\begin{figure} \center
\vspace {-0.5cm}
\includegraphics [width=60.52mm,height=68mm]{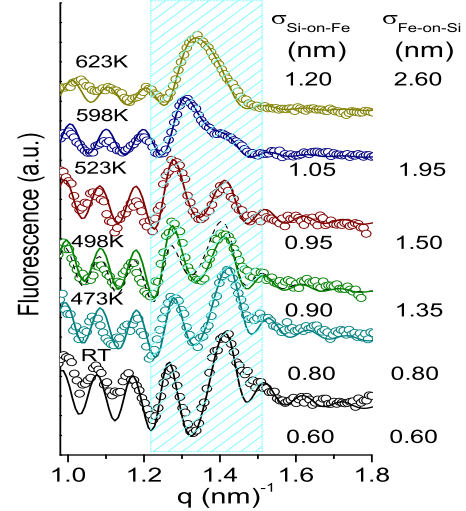}
\vspace {-0.3cm} \caption{\label{fig:fig2} Fe-fluorescence of
SW$_{-}$ML after various stages of annealing. The continuous
curves represent the best fit to the experimental data.The shaded
area represents the region around Bragg peak where x ray standing
waves are formed. The dashed curve represents the best fit to the
experimental data of 498K annealed sample with the roughnesses of
the two interfaces of Fe taken to be equal.}
\end{figure}

SW$_{-}$ML was isochronally annealed at 473\,K, 498\,K, 523\,K,
598\,K and 623\,K for 1\,h each.  Figure~\ref{fig:fig1} gives some
representative XRR of SW$_{-}$ML after various stages of
annealing. The XRR of this multilayer is dominated by that of W/Si
multilayer mirror, with only small modulation in the $q$ region
below the Bragg peak attributable to the top Si/Fe/Si trilayer.
Figure~\ref{fig:fig2} shows Fe-fluorescence data of SW$_{-}$ML
measured simultaneously with XRR.  The region around the Bragg
peak in which x ray standing waves are generated, is highlighted
in the figure by shaded area.  One may note that in the
as-deposited sample Fe-fluorescence exhibits two well defined
peaks in the region where x ray standing waves are generated.  The
origin of these two peaks can be understood from the contour plot
of x ray intensity (inset of Fig~\ref{fig:fig1}).  At the center
of the Bragg peak, Fe layer lies roughly midway between the two
antinodes.  However, as one moves away on either side of the Bragg
peak, the antinodes get shifted resulting in partial overlapping
of one of the antinodes with Fe layer, giving rise to a peak in
Fe-fluorescence. The peak around $q = 1.27\,nm^{-1}$ corresponds
to a situation where one of the antinodes partially overlaps with
Fe-on-Si interface while the peak at $q = 1.41\,nm^{-1}$ occurs as
a result of partial overlap of an antinode at Si-on-Fe interface.
With thermal annealing both these peaks get broadened and their
intensities get modified. However, changes occuring in the two
peaks are quite different, suggesting that the two interfaces get
modified differently with thermal annealing. Simultaneous fitting
of XRR and Fe-fluorescence data has been done using Parratt\'{}s
formalism~\cite{Parratt:PR:1954}.  The average roughness of the
interfaces of W/Si multilayer $\sigma_{W/Si}$, and the roughnesses
of the two interfaces of Fe, $\sigma_{Fe-on-Si}$ and
$\sigma_{Si-on-Fe}$ were the only parameters which were varied as
a function of annealing temperature.  While the XRR data is mainly
sensitive to the changes in $\sigma_{W/Si}$, the Fe-fluorescence
is affected by the changes in the structure of the interfaces of
Fe layer. One finds that with thermal annealing up to 623\,K,
there are only minor changes in the W/Si multilayer with the
interface roughness going from 0.6\,nm to 0.85\,nm.  The
roughnesses of the Fe-on-Si and Si-on-Fe interfaces as obtained
from the fitting of the fluorescence  and XRR data are shown in
Fig.~\ref{fig:fig2}.  One may note that the roughness of Fe-on-Si
interface increases at a much faster rate as compared to that of
Si-on-Fe interface.  It may be mentioned that for the thermal
annealing upto 473\,K, the width of Fe concentration profile is
much smaller than the separation between two antinodes. Therefore,
from the fluorescence data it is difficult to estimate the
roughnesses of the two interfaces separately.  However, at 498K
and above roughnesses of the two interfaces can be determined
individually with good reliability. In order to demonstrate this
point, the fitting of 498\,K data obtained by taking the
roughnesses of the two interfaces to be equal is also shown in
Fig.~\ref{fig:fig2} (dashed curve). The best fitted curve clearly
deviates from the experimental data.

\begin{figure} \center
\vspace {-0.5cm}
\includegraphics [width=62mm,height=55mm]{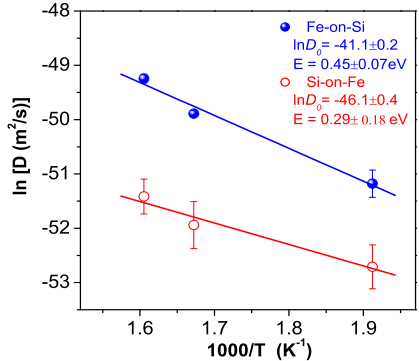}
\vspace {-0.4cm} \caption{\label{fig:fig3} Arrhenius plot of Fe
diffusivity at the two interfaces.}
\end{figure}

The variation in the roughnesses of the two interfaces with the
thermal annealing has been used to estimate the diffusivity of Fe
at the two interfaces using the relation:
$D(T_{2})=[{\sigma^{2}(T_{2})-\sigma^{2}(T_{1})}]/ {2t}$,where
D$(T_{2})$ is the diffusion coefficient at temperature $T_{2}$,
$\sigma(T_{1})$ and $\sigma (T_{2})$ are the roughnesses of an
interface before and after annealing at temperature $T_{2}$ for
time $t$. Figure.~\ref{fig:fig3} gives the Arrhenius plot of $lnD$
versus $1/T$ at the two interfaces of Fe. One finds that there is
a significant difference in Fe diffusivities at the two
interfaces. A linear fit to the experimental data as expected from
the Arrhenius temperature dependence of the diffusivity
$D(T)=D_{0} exp (-E/K_{B}T)$, yields the activation energy E as
well as the pre-exponent $D_{0}$ for the diffusion. The
experimentally obtained values are shown in Fig.~\ref{fig:fig3}.
One finds that the activation energies for diffusion at the two
interfaces are comparable within experimental errors.  However,
the pre-exponent $D_{0}$ is significantly high for Fe-on-Si
interface as compared to that for Si-on-Fe interface, resulting in
a significantly high diffusivity at Fe-on-Si interface in the
temperature range studied in the present experiment.

At first glance the above results appear to be counter intuitive
as both interfaces ought to be identical, having Fe layer on one
side and Si layers on the other.  In order to understand the
possible reason for this difference in the diffusivity at the two
interfaces, a detailed study of the structure of the two
interfaces has been done using CEMS on the samples ML1 and ML2. As
discussed in the experimental section CEMS of ML1 is sensitive to
Fe-on-Si interface while that of ML2 is sensitive to Si-on-Fe
interface.

\begin{figure} \center
\vspace {-0.5cm}
\includegraphics [width=75.625mm,height=55mm]{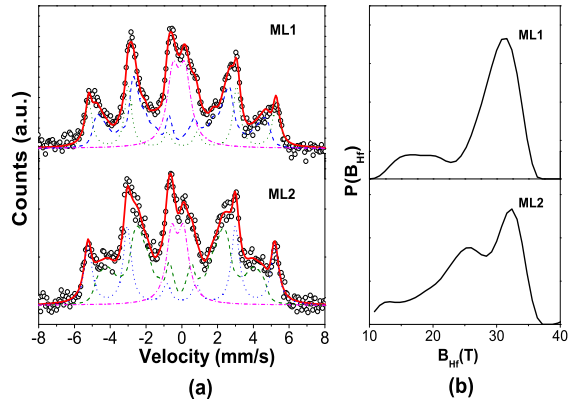}
\vspace {-0.4cm} \caption{\label{fig:fig4} (a) Conversion electron
M\"{o}ssbauer spectra of as-deposited ML1 and ML2,(b) The
corresponding hyperfine field distribution of the broad magnetic
sextet.}
\end{figure}

\begin{table}
\caption{\label{tab:table1} Results of fitting of M\"{o}ssbauer
spectra of ML1 and ML2 as described in the text. $\delta$,
$\Delta$ and B$_{hf}$ represent isomer shift, quadrupole splitting
and average hyperfine field respectively. Isomer shift $\delta$ is
with respect to $\alpha$-Fe.}
\begin{ruledtabular}
\begin{tabular}{|cc|c|c|}
Sample& &ML1&ML2 \\  \hline
Sharp Sextet&B$_{hf} (T)$&31.34$\pm$0.08&32.32$\pm$0.06\\
&Area (\%) &28$\pm$6&32$\pm$5\\ \hline
Broad Sextet&B$_{hf} (T)$&26.11$\pm$0.32&25.74$\pm$0.21\\
&Area (\%) &46$\pm$6&50$\pm$3\\ \hline
&$\delta$ (mm/s)&0.16$\pm$0.01&0.24$\pm$0.01\\
Doublet&$\Delta$ (mm/s)&0.68$\pm$0.02&0.65$\pm$0.03\\
&Area (\%) &26$\pm$3&18$\pm$3\\
\end{tabular}
\end{ruledtabular}
\end{table}

\begin{figure} \center
\vspace {-0.5cm}
\includegraphics [width=53.2mm,height=40mm]{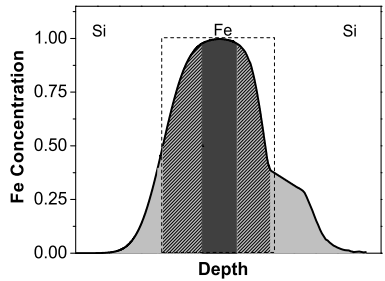}
\vspace {-0.4cm} \caption{\label{fig:fig5} Schematic diagram of
concentration profile of Fe layer. The regions corresponding to
sharp sextet, broad sextet and doublet are shown in different
shades. The dashed rectangle shows the position of ideal
interfaces.}
\end{figure}

Figure~\ref{fig:fig4}(a) shows M\"{o}ssbauer spectra of ML1 and
ML2 in as-deposited state.  The spectra of both the specimens are
fitted with three overlapping components: (1) a sharp sextet with
hyperfine field about 33\,T, (2) a broad magnetic component,
having a distribution of hyperfine fields and (3) a non-magnetic
doublet.  The sharp sextet represents the bulk of $\alpha$-Fe,
while the broad magnetic component and the doublet represent the
Fe atoms in the interfacial region.  Further it is known that in
Fe-Si alloy, if iron concentration is less than 50\%, it becomes
non-magnetic~\cite{Walterfang:PRB:2006}.  Therefore, the area
under the doublet represents the fraction of Fe atoms in the
interfacial region having iron concentration less than or equal to
50\%. Results of fitting are given in table (1). One may note that
the area under the sharp sextet in both the specimens is about
30\%, within experimental errors.  However, the relative areas of
broad magnetic component and the doublet as well as the shape of
the hyperfine field distribution of broad magnetic component
[Fig.~\ref{fig:fig4}(b)] are very different in the two specimens.
This suggests that the structure of the two interfaces namely
Si-on-Fe and Fe-on-Si interface may be different.  A rough
estimate of the interface roughness as obtained from the width of
the intermixed region comes out to be 0.78 nm, which is in
agreement with that obtained from x ray measurements.  From table
(1), one finds that at Si-on-Fe interface, the fraction of Fe
atoms in the doublet is 18\%, which is about 23\% of the total Fe
atoms in the intermixed region.  This area fraction agrees very
well with that expected for an error function concentration
profile [Fig.~\ref{fig:fig5}].  Thus, M\"{o}ssbauer measurements
suggest that the concentration profile at Si-on-Fe interface is an
error function.

From M\"{o}ssbauer spectrum of specimen ML1 one finds that in this
spectrum, the area of doublet is 27\%, which is significantly more
than what one expects for an error function concentration profile.
Further a comparison of hyperfine field distributions at the two
interfaces shows that at Si-on-Fe interface, there is a continuous
distribution of hyperfine field values starting from 32\,T down to
10\,T while at Fe-on-Si interface, low field components are
missing.  The above two differences can be understood, if one
assumes the concentration profile at Fe-on-Si interface to be as
shown schematically in Fig.~\ref{fig:fig5}, where the area from
the low field magnetic component gets transferred to the
non-magnetic component. This suggests that at Fe-on-Si interface,
there is an interlayer of Fe$_{1-x}$Si$_{x}$, resulting in a
plateau in a concentration profile. The isomer shift and
quadrupole values of the doublet of ML1 match very well with those
of FeSi$_{2}$, suggesting that the composition of interlayer is
FeSi$_{2}$~\cite{Desimoni:PRB:1996}. Thus, at Fe-on-Si interface
diffusion occurs via the FeSi$_{2}$ interlayer, while at Si-on-Fe
interface diffusion occurs via bcc Fe (Si) phase. In literature
the diffusivity data for self-diffusion in stiochometric D$O_{3}$
phase as well as intermetallic compounds of FeSi are given.
However, it is not meaningful to compare the diffusivities
obtained in the present work at the two interfaces with bulk
diffusivities, since it is known that the multilayer diffusivities
at the interfaces are very different because of high concentration
of defects and possible concentration
gradient~\cite{Wang:PRB:1999}.

The difference in the structure of the two interfaces can be
understood in terms of the difference in the surface free energies
of Fe(2.9$Jm^{-2}$) and Si (1.2$Jm^{-2}$)~\cite{Himpsel:MNS:1998}.
During the deposition of Fe on Si, the surface free energy of Si
being lower, Si atoms try to move to the surface guided by the
chemical driving force. This would lead to a stronger mixing at
the interface and a possible formation of FeSi$_{2}$ compound.  On
the other hand, during the deposition of Si on Fe, no such
chemical driving force exists, therefore the intermixing at
Si-on-Fe interface would take place as a result of random thermal
motions only and hence concentration profile is expected to be an
error function.

In conclusion, x ray standing wave technique has been used to get
concentration profile of Fe layer in Si/Fe/Si trilayer.  The
precision of this technique is sufficient to differentiate between
the two interfaces of the Fe layer.  This allows one to study
interdiffusion at the two interfaces namely Fe-on-Si and Si-on-Fe
independently.  Interestingly the diffusivities at the two
interfaces are significantly different.  This seemingly
counter-intuitive result can be understood in terms of a
difference in the structure of the two interfaces in the
as-deposited film itself.  CEMS measurements show that while at
Si-on-Fe interface, the Fe concentration profile is an error
function, at Fe-on-Si interface an interlayer exists with
approximate composition of FeSi$_{2}$.  This difference in the
structure of the two interfaces is the cause of different
diffusivities at the two interfaces.  Besides being of fundamental
importance in understanding the interfacial diffusion at nanometer
scale, present results have important implications on the use of
controlled thermal annealing for tailoring the properties of
multilayers for a wide variety of applications.

Acknowledgement: Partial support from the Indo-French Center for
Promotion of Advanced Research is acknowledged.

\bibliography {Fe_si}

\end{document}